\documentclass
[twocolumn,secnumarabic,amsmath,amssymb,showpacs,nobibnotes,aps, prl]{revtex4-1}

\usepackage{amssymb}

\usepackage{graphicx}
\usepackage{dcolumn}
\usepackage{bm}
\usepackage{color}

\begin{document}
\title{Spontaneous spin-up induced by turbulence-driven topological transition of orbits in a collisionless tokamak plasma}
\author{Shaojie Wang}
\email{wangsj@ustc.edu.cn}
\affiliation{Department of Engineering and Applied Physics, University of Science and Technology of China, Hefei, 230026, China}
\date{\today}

\begin{abstract}
Spontaneous spin-up are widely observed in tokamak plasmas, which is crucially important for plasma confinement. A kinetic theory is proposed to show that a toroidal rotation of core plasma is induced by the topological transition of orbits driven by turbulent diffusion in a collisionless tokamak plasma. The theoretical prediction agrees well with the well-known Rice-scaling of intrinsic core plasma flow. This new theory predicts an intrinsic co-current core parallel flow of $\sim 100km/s$ in the International Thermonuclear Experimental Reactor.
\end{abstract}


\maketitle

\section{Introduction}

Spontaneous spin-up is widely observed, from particles to galaxies, for examples, the well-known giant zonal belt on the Jupiter \cite{PerezIca12}. In a tokamak, a magnetic fusion torus in the shape of a donut, spontaneous toroidal spin-up of core plasma is routinely observed after the Low-confinement to High-confinement (L-H) transition \cite{LeePRL03,RiceNF07,RicePPCF08,SolomonNF09,PeetersNF11,AngioniNF12,IdaPRL13}. To understand the physical mechanism of this so-called intrinsic flow, various theoretical models have been proposed: the effect of the residual Reynolds stress \cite{DiamondPoP08,DiamondNF13} on the momentum redistribution \cite{GriersonPRL17}, the effect of turbulence intensity gradient \cite{StoltzfusPRL12} and the effect of thermal ion orbit loss \cite{deGrassieNF12,StacyNF13,PanNF14,BoedoPoP16} on the boundary flow, the effect of Coriolis force on the momentum pinch \cite{PeetersPRL07,HahmPoP07}, and the turbulent acceleration \cite{WangPRL13,GarbetPoP13}. However, it is still an open issue to predict the net core flow in ITER. In this paper, we show that a spontaneous toroidal flow can be induced by the topological transition of ion orbit in a collisionless turbulent tokamak plasma.

\section{Basic Equations}

The equilibrium magnetic field of a tokamak is written as $\bm B =I\left(r\right)\nabla \zeta +\nabla \zeta \times \nabla \psi \left(r\right)$, with $I=R B_T$, $B_T$ the toroidal magnetic field, $R$ the major radius. The poloidal magnetic flux is $\psi\left(r\right)$, with $r$ essentially the minor radius of the torus. The poloidal magnetic field is given by $B_{P}=\psi'/R$, with the prime denoting the derivative with respect to $r$. $\zeta$ is the toroidal angle. In this paper, we shall consider a large-aspect-ratio ($\epsilon\equiv r/R\ll 1$,) up-down symmetric tokamak.

The ensemble averaged distribution function is $f\left(\psi,\theta,\mu,w,t\right)$, with $\theta$ the poloidal angle, $\mu=v_{\perp}^{2}/2B$ the magnetic moment, and $w=v_{\|}^{2}/2+\mu B\left(\psi,\theta\right)$ the energy of the particle; $v_{\perp}$ and $v_{\|}$ are the velocity components perpendicular and parallel to the magnetic field, respectively. Since the momentum of plasma is mainly carried by the ions, we consider only the ion dynamics in the following.

We begin with the transport equation,
\begin{equation}
\partial_t f +\mathcal{L}\left(f\right)+\mathcal{T}\left(f\right)=S,\label{eq:Trans}
\end{equation}
where the orbiting term is given by
\begin{equation}
\mathcal{L}\left(f\right)
\equiv v_{\|}\mathbf{b}\cdot \nabla f +\mathbf{V}_{d}\cdot \nabla f, \label{eq:L}
\end{equation}
with $\mathbf{b}=\mathbf{B}/B$. The guiding-center drift (GC) velocity is given by the Alfven approximation \cite{HintonRMP76},
$\mathbf{V}_{d}=-v_{\|}\mathbf{b}\times\nabla \rho_{\|}$,
with $\rho_{\|}=mv_{\|}/eB$; $e$ and $m$ are the charge and mass of the particle, respectively. The Jacobian of the phase space is given by
$1/J=|v_{\|}|\mathbf{b}\cdot \nabla \theta/\left(2\pi \right)$.

The radial transport term including the effect of turbulence \cite{KadomtsevBOOK70,DupreePF66,StoltzfusPRL12,WangPoP16a} is given by
\begin{equation}
\mathcal{T}\left(f\right)=-\frac{1}{J}\partial_{\psi}\left(J\mathcal{D}\partial_{\psi}f\right), \label{eq:Turbu}
\end{equation}
where $\mathcal{D}=\left(\psi'\right)^2$D, with $D$ the usual radial diffusivity due to the turbulence.

To concentrate on the intrinsic rotation, we shall assume that the source term $S=S\left(\psi\right)$ contains the particle source, $\int d^3\mathbf{v}S\left(\psi\right)\neq0$, and the energy source, $\int d^3\mathbf{v} wS\left(\psi\right)\neq0$, but it does not contains the momentum source, $\int d^3\mathbf{v} v_{\|}S\left(\psi\right)=0$. Note that
\begin{equation}
d^3\mathbf{v}=\sum_{\sigma}\frac{2\pi B}{|v_{\|}|}d\mu dw,
\end{equation}
where $\sigma$ is the sign of $v_{\|}$.
We shall also assume that the turbulent diffusivity $\mathcal{D}\left(\psi\right)$ is independent of $\sigma$; for the electrostatic turbulence, diffusion is induced by the fluctuating $\mathbf{E}\times\mathbf{B}$ drift, whose dependence on the velocity is weak.

Following Ref. \onlinecite{StoltzfusPRL12}, we ignored the collision term, since we are concentrating here on the H-mode plasma.

The GC orbit in the equilibrium fields is well-defined by the three constants of motion (COMs), the magnetic moment, the particle energy, and the toroidal canonical angular momentum ($-eP$), which is given by
\begin{equation}
P=\psi-\rho_{\|}\left(\psi,\theta,\mu,w,\sigma\right)I,\label{eq:P}
\end{equation}
Using the three COMs, one finds \cite{PutvinskiiBook93}
\begin{equation}
\frac{\left(R-0.5R_{b}\right)^2}{\left(0.5R_{b}\right)^2}
-\frac{\left(\psi-P\right)^2}{\left(0.5R_{b}mv/e\right)^2}=1,\label{eq:orbit}
\end{equation}
with $R_{b}=\mu I/\left(0.5mv^2\right)$. This equation defines a hyperbola in the $R-\psi$ plane, with its tip located at $\left(R_{b},P\right)$.

The equilibrium poloidal magnetic flux is given by
\begin{equation}
\psi=\psi\left(R, Z\right), \label{eq:eq}
\end{equation}
where $(R,Z,\zeta)$ is the cylindrical coordinates.

The GC orbit in the $R-Z$ plane (the minor cross-section of the torus) is determined by Eqs. (\ref{eq:orbit}, \ref{eq:eq}). When the tip $\left(R_{b},P\right)$ locates inside $\psi=\psi_{E}\left(R\right)\equiv \psi\left(R,Z=0\right)$, the particle is trapped and its orbit in the $R-Z$ plane is a banana orbit; otherwise the particle is passing and its orbit in the $R-Z$ plane is approximately a circle. For a passing orbit, one needs $\sigma$, in addition of the three COMs, to completely determine the orbit. GC orbits in the $R-\psi$ plane in terms of the COMs are schematically shown in Fig. 1. When including an electrostatic potential $\phi\left(\psi\right)$, the above discussions on the GC orbit are slightly modified, with the tip $\left(R_{b},P\right)$ slightly shifted, and this shift is similar for all particles; for further details, we refer the readers to Ref. \onlinecite{PutvinskiiBook93}.

\begin{figure}
  \includegraphics[width=6.0cm]{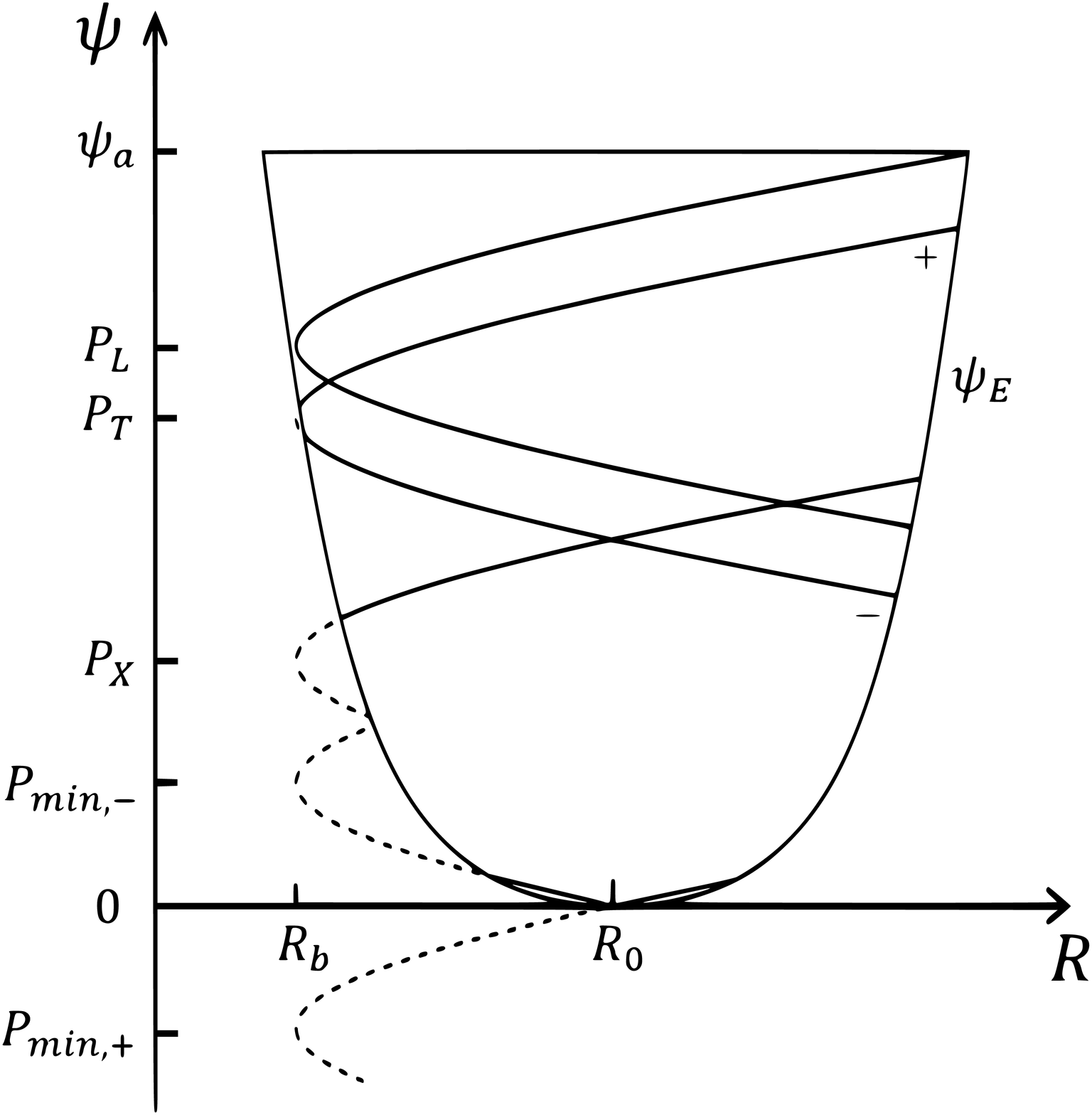}\\
  \caption{GC orbits with given $(\mu,w)$ in $R-\psi$ space. A trapped/passing particle has 2/1 (a topological number) cross points with the right branch of the curve $\psi_E$. Diffusion of a co-passing orbit from $(P_{min},+)$ to $(P_{T},+)$ or a counter-passing orbit from $(P_{min},-)$ to $(P_T,-)$ induces a topological transition at the trapping-passing boundary, from $(P_{T},+/-)$ to $(P_{L},T)$.} \label{fig1}
\end{figure}

In a large aspect-ratio tokamak with concentric circular magnetic flux surface, if one introduces the constant $-q$ approximation, one finds
\begin{equation}
\psi\left(R, Z\right)=\frac{B_{0}}{2q}\left[\left(R-R_{0}\right)^2+Z^2\right],
\end{equation}
where $q$ is the safety factor, and the subscript $0$ means the corresponding value is evaluated at the magnetic axis. By definition, $\psi=0$ at the magnetic axis; note that this equation defines a parabola in the $R-\psi$ plane with $\psi=\psi_{E}\left(R\right)$.

Clearly, given $\left(\mu,w,\sigma\right)$, the minimum value of $P$ is given by
\begin{equation}
P_{min,\sigma}=-I\rho_{\|,0}, \label{eq:Pmin}
\end{equation}
where $\rho_{\|,0}=\frac{m v_{\|,0} } {eB_{0}}$, with $v_{\|,0}=\sigma\sqrt{2\left(w-\mu B_{0}\right)}$. Note that
\begin{equation}
\Delta P\left(\mu,w\right)\equiv P_{min,-}-P_{min,+}=2I|\rho_{\|,0}|. \label{eq:Delta-P}
\end{equation}

Let $\psi_O\left(P,\mu,w\right)$ be the outermost radial position that a particle labeled by $(\mu,w,\sigma)$ launched from $(\psi,\theta)$ can reach at $\theta=\theta_{O}$, where $\theta_{O}=0$ for trapped particles and co-passing particles, and $\theta_{O}=\pi$ for counter-passing particles. Using the COMs, one finds
\begin{equation}
\psi_{O}-\rho_{\|,O}I=P, \label{eq:B1}
\end{equation}
\begin{equation}
\frac{1}{2}v_{\|,O}^2+\mu B\left(\psi,\theta_{O}\right)=w. \label{eq:B2}
\end{equation}
Eq. (\ref{eq:B1}) and Eq. (\ref{eq:B2}) define a function $\psi_{L}$ through $\psi_{L}\left(\psi_O,\theta,\mu,w\right)=\psi$; therefore, the orbit loss condition is given by
\begin{equation}
\psi\geq \psi_{L}\left(\psi_O,\theta,\mu,w,\sigma\right)_{\psi_{O}=\psi_{a}}, \label{eq:LC}
\end{equation}
with $\psi_a$ the poloidal magnetic flux at the boundary of the torus. Note that this is a simple orbit-loss model, for a specific machine, the orbit-loss condition may be modified, however, the method developed here can be straightforwardly extended.

Clearly, a boundary condition
\begin{equation}
\left[f\right]_{\psi\geq\psi_{L}}=0,
\end{equation}
should be associated with Eq. (\ref{eq:Trans}).

\section{Orbital Averaged Transport Equation and Topological transition of Orbits}

\subsection{Orbital Average}

To solve Eq. (\ref{eq:Trans}), we transform to the coordinate system, $\left(P, \mu,w,\theta\right)$. In this new coordinate system,

\begin{equation}
\mathcal{L}=\dot{\theta}\partial_{\theta},
\end{equation}
with the poloidal angular velocity given by
\begin{equation}
\dot{\theta}=\left(v_{\|}\mathbf{b}+\mathbf{V}_{d}\right)\cdot \nabla \theta =v_{\|}\mathbf{b}\cdot\nabla\theta\partial_{\psi}P\left(\psi,\theta,\mu,w\right). \label{eq:vp}
\end{equation}

Using Eq. (\ref{eq:vp}), one finds
\begin{equation}
\mathcal{T}\left(f\right)=-\frac{1}{\mathcal{J}}\partial_{P}\left(\mathcal{J}\mathcal{D}\partial_{P}f\right), \label{eq:Turbu-1}
\end{equation}
with the Jacobian of the new coordinate system given by
$1/\mathcal{J}=|\dot{\theta}|/\left(2\pi \right)$,
which is found by using Eq. (\ref{eq:vp}).

Note that $\partial_t f \sim S\sim\mathcal{T}\sim f/\tau_{E}$, with $\tau_{E}$ the confinement (turbulent diffusion) time; $\mathcal{L}\sim 1/\tau_{\theta}$, with
$\tau_{\theta}=\oint d\theta /\dot{\theta}$, the bounce time for trapped particles or the transit time for passing particles. For passing particles, $\oint d\theta=\int _{0}^{2\pi}d\theta$; for trapped particles, $\oint d\theta=\int _{-\theta_{b}}^{+\theta_{b}}d\theta+\int _{+\theta_{b}}^{-\theta_{b}}d\theta$, with $\pm\theta_{b}$ the bounce angle. Note that $\sigma$ is a constant of motion for passing particles, while it is not for trapped particles.

To proceed, we assume that
\begin{equation}
\delta\equiv \tau_{\theta}/\tau_{E}\ll 1. \label{eq:ordering}
\end{equation}

Expanding the distribution function with respect to $\delta$, we found $f=F+\delta f$.
To $\mathcal{O}\left(\delta^0\right)$, one finds
$\mathcal{L}\left(F\right)=0$,
which demonstrates that the lowest order solution is a constant of motion,
\begin{equation}
F=F\left(P,\mu,w\right). \label{eq:F0}
\end{equation}

To the next order, one finds
\begin{equation}
\mathcal{L}\left(\delta f\right)+\partial_t F +\mathcal{T}\left(F\right)=S.
\end{equation}

Orbital-averaging this equation, one finds
\begin{equation}
\partial_{t} F -\frac{1}{\tau_{\theta}}\partial_{P}\left[\tau_{\theta} \mathcal{D}\left(\overline{\psi}\right)\partial_{P} F\right]=S\left(\overline{\psi}\right). \label{eq:Trans-Fa}
\end{equation}
The orbital-averaging operator, which is an annihilator of $\mathcal{L}$, is defined by
\begin{equation}
\overline{ A}=\frac{1}{\tau_{\theta}}\oint \frac{d\theta}{\dot{\theta}} A.
\end{equation}
Using Eq. (\ref{eq:P}), one finds
\begin{equation}
\overline{\psi}-\psi=\left(\overline{\rho_{\|}I}-\rho_{\|}I\right);
\end{equation}
clearly, $\overline{\psi}=\overline{\psi}_{\sigma}\left(P,\mu,w\right)$. Note that Eq. (\ref{eq:Trans-Fa}) includes the finite-banana-width effects, since the orbital-average retains the effects of radial drift in Eq. (\ref{eq:L}).

In terms of the COMs, the orbit loss condition is
\begin{equation}
P\geq P_{L}\left(\mu,w,\sigma\right),
\end{equation}
which specifies the outer boundary condition of Eq. (\ref{eq:Trans-Fa}),
\begin{equation}
\left[F\right]_{P\geq P_{L}}=0.
\end{equation}

The inner boundary condition for Eq. (\ref{eq:Trans-Fa}) is given by
\begin{equation}
\left[\tau_{\theta} \mathcal{D}\left(\overline{\psi}\right)\partial_{P} F\right]_{P=P_{min,\sigma}}=0,
\end{equation}
which is similar to the natural boundary condition used previously in neoclassical transport theory \cite{ZaitsevPFB93}.

Note that the phase space element in terms of $\left(P, \mu, w\right)$ is $2\pi dP\times2\pi d\mu\times \tau_{\theta}dw$; $\tau_{\theta}$ is the Jacobian of the COM space $\left(P,\mu,w\right)$. The number of particles $N$ is given by
\begin{equation}
dN=\sum_{\sigma}4\pi^2\tau_{\theta}dP d\mu dw F, \label{eq:Jp0}
\end{equation}
with $\sigma=+,-$ for passing particles, and $\sigma=T$ for trapped particles.

Before further discussions, we make general comments on Eq. (\ref{eq:Trans-Fa}). The basic concept here is that the turbulent diffusion of ions is much slower than the orbiting process, therefore it is essentially the GC drift center rather than the particle itself that is diffused by the turbulence. This has been numerically confirmed by examining the ion orbits in a typical Ion-Temperature-Gradient-driven turbulence, which are shown in Fig. 2 and Fig. 3; the fluctuating field is given by the nonlinear gyrokinetic simulation \cite{XuPoP17}. Note that the equilibrium orbits, which are determined by the COMs, still can be read from Fig. 2 and Fig. 3, during the turbulent diffusion process.

\begin{figure}
  \includegraphics[width=6.0cm]{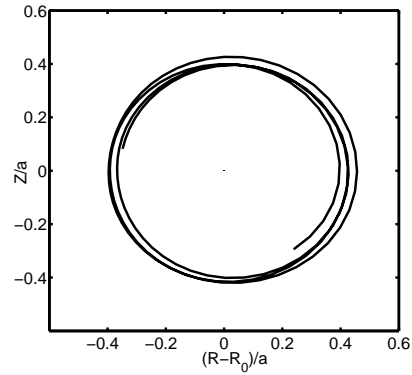}\\
  \caption{Typical passing ion orbit in the ITG turbulence.} \label{fig2}
\end{figure}

\begin{figure}
  \includegraphics[width=6.0cm]{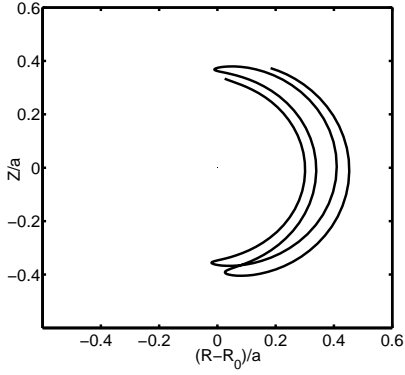}\\
  \caption{Typical trapped ion orbit in the ITG turbulence.} \label{fig3}
\end{figure}

\subsection{Topological Transition of Orbits}
When the passing particles, which carry the parallel momentum in the core, diffuse radially outward, both the co-passing orbit and the counter-passing orbit may undergo a topological transition to a trapped orbit, as is schematically shown in Fig. 1. This fundamental concept has been confirmed by numerical simulation of the ITG turbulence; typical orbit transition induced by the turbulent diffusion is shown in Fig. 4.

\begin{figure}
  \includegraphics[width=6.0cm]{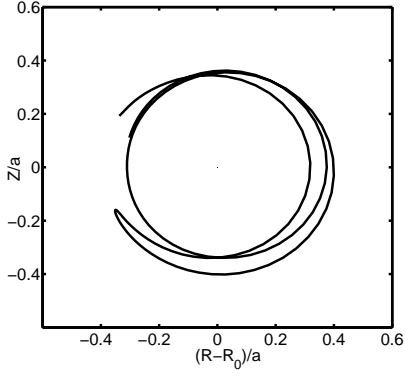}\\
  \caption{Typical orbit transition induced by the turbulent diffusion.} \label{fig4}
\end{figure}

The spontaneous co-current spin-up of core plasma after L-H transition of a tokamak can be understood without the final solution. After L-H transition, the plasma enters the collisionless state, and particles in the boundary region can complete their trapping (banana) orbit, which dominates the orbit loss; when a co/counter-passing particle diffuses radially outward, it may change to a trapped orbit through the topological transition of orbit at $P=P_{T}$; therefore, the co-moving particles have larger confinement region [$(P_{min,+},P_{T})$ in Fig. 1] than the counter-moving particles [$(P_{min,-},P_{T})$ in Fig. 1]; this $\sigma-$ asymmetry induced by the topological transition clearly generates a co-current flow in the core region.
Therefore, one may estimate the anisotropy in the core region as follows. Using the fact that the confinement regions of co/counter-passing particles with given $\left(\mu,w\right)$, which are illustrated in Fig. 1, are $\left(P_{L}-P_{mim,+}\right)$ and $\left(P_{L}-P_{mim,-}\right)$, respectively, one finds the difference between the confinement regions is $\Delta\psi_{B}=\Delta P=2|\rho_{\|,0}|I$, and therefore

\begin{equation}
F_{+}-F_{-}\sim S \frac{\psi_{a}^2}{\mathcal{D}}\frac{\Delta\psi_{B}}{\psi_{a}}, \label{eq:anis}
\end{equation}
with $F_{+}\sim S \psi_{a}^{2}/\mathcal{D}$.

Note here that this spontaneous spin-up of core plasma in a turbulent tokamak induced by the topological transition of orbits is different from the topological phase transition in condensed matter physics \cite{KosterlitzJPC72,KosterlitzJPC73}.

\section{Solution to the Orbital-Averaged Transport Equation}
Eq. (\ref{eq:Trans-Fa}) should be solved for $F_{\sigma}$, with $\sigma=+,-,T$ for co-current passing, counter-current passing, and trapped particles, respectively; therefore the phase-space is cut into three sub-domains. Note that with given $\left(\mu,w\right)$, when $P<P_{T}$, the particle is passing; when $P>P_{T}$, the particle is trapped. Following Ref. \onlinecite{ZaitsevPFB93}, we introduce the connection formulas in the Trapping-Passing Boundary (TPB) where $P=P_{T}$:
\begin{equation}
\left[F_{+}\right]_{TPB}=\left[F_{-}\right]_{TPB}=\left[F_{T}\right]_{TPB}; \label{eq:connection-dist}
\end{equation}

\begin{equation}
\left[\tau_{\theta} \mathcal{D}\partial_{P} F_{+}\right]_{TPB}+\left[\tau_{\theta} \mathcal{D}\partial_{P} F_{-}\right]_{TPB}=\left[\tau_{\theta} \mathcal{D}\partial_{P} F_{T}\right]_{TPB}. \label{eq:connection-flux}
\end{equation}

To solve Eq. (\ref{eq:Trans-Fa}), it is useful to note that the dependence of $P_{min,\sigma}\left(\mu,w\right)$ on $\sigma$ is strong [see, Eq. (\ref{eq:Pmin})], while the dependence of $\overline{\psi}_{\sigma}\left(P_{min, \sigma},\mu,w\right)$ on $\sigma$ is weak. This can be understood as follows. Define
\begin{equation}
P_{X}=P-\Delta P, \label{eq:PX}
\end{equation}
for co-passing particles. It is not hard to understand that for well-passing particles,
\begin{equation}
\overline{\psi}_{\sigma=+}\left(P_{X},\mu,w\right)\approx\overline{\psi}_{\sigma=-}\left(P,\mu,w\right). \label{eq:PX-}
\end{equation}
This can also be understood in an alternative way. It is well-known that a passing orbit deviates from the magnetic flux surface with a horizontal shift $\Delta R=q\rho_{\|}$, and a passing orbit labeled by $P_{min,\sigma}$ passes through the magnetic axis; therefore, $\overline{\psi}_{\sigma}\left(P_{min, \sigma},\mu,w\right)\simeq \psi\left(q_{0}|\rho_{\|,0}|\right)$; see also Fig. 1.

Before proceeding, we shall assume that the source term is zero in the boundary region so that
\begin{equation}
S\left(\overline{\psi}\right)=0,
\end{equation}
when $P_{X}\geq P_{L}$ for $\sigma=+$, or $P\geq P_{L}$ for $\sigma=-$. This ensures that there is no source directly input into the loss-cone.

The steady-state solution of Eq. (\ref{eq:Trans-Fa}) is now readily found. For given $\left(\mu,w\right)$, the flux, $\Gamma_{\sigma}=-\mathcal{D}\left(\overline{\psi}\right)\partial_{P}F_{\sigma}$, is given as follows.
For passing particles ($P<P_{T}$),
\begin{equation}
\Gamma_{\pm}=\frac{1}{\tau_{\theta}}
\int_{P_{min,\pm}}^{P}dP'
\tau_{\theta} S\left[\overline{\psi}\left(P',\mu,w\right)\right]; \label{eq:solution-flux-passing}
\end{equation}
for trapped particles ($P>P_{T}$),
\begin{equation}
\Gamma_{T}=\frac{1}{\tau_{\theta}}
\int_{P_{T}}^{P}dP'
\tau_{\theta} S\left[\overline{\psi}\left(P',\mu,w\right)\right]-\left[ \mathcal{D}\partial_{P} F_{T}\right]_{TPB}. \label{eq:solution-flux-trapped}
\end{equation}

The steady-state distribution function is given as follows.
For trapped particles,
\begin{equation}
F_{T}=\int_{P_{L}}^{P}  dP'
 \partial_{P}F_{T}\left[\bar{\psi}\left(P',\mu,w\right),\mu,w\right]; \label{eq:solution-F-trapped}
\end{equation}
for barely-passing particles which satisfy $P_{T}<P_{L}$,
\begin{equation}
F_{\pm}=\int_{P_{T}}^{P}  dP'
 \partial_{P}F_{\pm}\left[\bar{\psi}\left(P',\mu,w\right),\mu,w\right]+\left[F_{\pm}\right]_{TPB}; \label{eq:solution-F-passing}
\end{equation}
for well-passing particles which satisfy $P_{T}>P_{L}$,
\begin{equation}
F_{\pm}=\int_{P_{L}}^{P}  dP'
 \partial_{P}F_{\pm}\left[\bar{\psi}\left(P',\mu,w\right),\mu,w\right]. \label{eq:solution-F-well-passing}
\end{equation}

\section{Spontaneous Core Plasma Toroidal Rotation Induced by the Orbit Transition}

Clearly, the anisotropy is weaker in the core region where $P<P_{L}$ than in the boundary region where $P\geq P_{L}$. The width of the boundary region is given by
$\psi_{a}-\psi_{B}\sim \Delta \psi_{B}\sim |\rho_{\|}I|$.
Since the orbit loss is dominated by the initially counter-moving trapped particles, it can be estimated that
\begin{equation}
-\partial_{\psi}T_{B}\sqrt{\epsilon}v_{th,B}I/\Omega=T_{B}, \label{eq:TB}
\end{equation}
with $\partial_{\psi}T_{B}$ the radial gradient of temperature in the boundary region, and $T=\frac{1}{2}mv_{th}^2$. The temperature $T$ is given by $p=nT=\int d^3\mathbf{v}\frac{1}{3}mv^2F$, with $n=\int d^3\mathbf{v}F$ the density, and $p$ the pressure. Note that $\Delta \psi_{B}\sim I\sqrt{\epsilon}v_{th,B}/\Omega$ is the half banana-width of the typical trapped particles launched from the typical orbit-loss boundary labeled by $\psi_{B}$, where the ion temperature is $T_{B}=\frac{1}{2}mv_{th}^2$. Note that $T_{B}$ is typically the pedestal temperature of the H-mode plasma. Eq. (\ref{eq:TB}) is found by using $-\partial_{\psi}T_{B}\Delta \psi_{B}=T_{B}-T\left(\psi_{a}\right)$, with $T\left(\psi_{a}\right)=0$.

Therefore, one finds the typical parallel velocity of the trapped particle at the orbit-loss boundary
\begin{equation}
\sqrt{\epsilon}v_{th,B}=-2\frac{a}{e}\partial_{\psi}T_{B}. \label{eq:LRW}
\end{equation}

The ion parallel flow, $u$, is defined by
\begin{equation}
nu=\left\langle \int d^3\mathbf{v}v_{\|}F_{\sigma}\left(\psi-\rho_{\|}I,\mu,w\right)\right\rangle.\label{eq:nu}
\end{equation}

The magnetic-flux-surface average operator is given by
\begin{equation}
\left\langle A\right\rangle=\frac {1}{\oint\frac {d\theta}{\mathbf{B}\cdot\nabla \theta}}\oint \frac {d\theta}{\mathbf{B}\cdot\nabla \theta}A.
\end{equation}

The parallel momentum in the core region can be estimated as follows. For the well-passing particles in the core, when they diffuse radially outward, they do not undergo an orbit transition before they leak out of the system at the boundary; however, the orbit-loss of well-passing particles defines nearly equal confinement domains for co-current and counter-current well-passing particles; therefore, the parallel momentum contained in Eq. (\ref{eq:solution-F-well-passing}) can be ignored.

For the barely-passing particles which undergo orbit transitions, its anisotropy is contained in Eq. (\ref{eq:solution-F-passing}); this anisotropy can be evaluated as follows.
Define $F_{d}$ with
\begin{equation}
F_{\sigma}=F_{d}+F_{\sigma,ped}, \label{eq:F-ped}
\end{equation}
\begin{equation}
F_{+,ped}=\int_{P_{T}}^{P_{X}} dP'
 \partial_{P}F_{+}\left[\bar{\psi}\left(P',\mu,w\right),\mu,w\right], \label{eq:df}
\end{equation}
and $F_{-,ped}=0$. Note that $P_{T}-P_{X}=\Delta P=P_{min,-}-P_{min,+}$. 

Clearly, the anisotropy contained in $F_{d}$ is weak, this can be understood by examining Eqs. (\ref{eq:PX}, \ref{eq:PX-}, \ref{eq:solution-flux-passing}, \ref{eq:solution-F-passing}). Therefore, the anisotropy due to the orbit-loss boundary condition is mainly contained in $F_{\sigma,ped}$, which can be taken as a "pedestal" of anisotropy schematically shown in Fig. 5.

\begin{figure}
  \includegraphics[width=6.0cm]{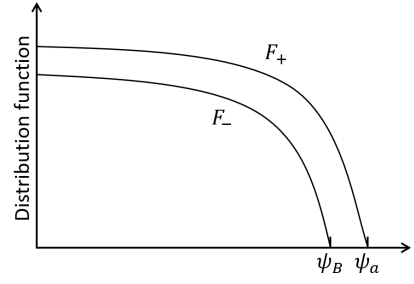}\\
  \caption{Pedestal of the distribution anisotropy.} \label{fig5}
\end{figure}

The core momentum due to the weak anisotropy contained in $F_{d}$ can be evaluated in a straightforward way by using the method previously developed \cite{RosenbluthPRL98,WangPoP17},
\begin{equation}
\left[nmu\right]_{d}=\left\langle \int d^3\mathbf{v} mv_{\|}F_{d}\right\rangle=-1.6\epsilon^{3/2}\frac{1}{\Omega_{p}}\left(-ne\partial_{r}\phi-\partial_{r}p\right), \label{eq:weak-a}
\end{equation}
where the radial electric field effect is retained; by using the ion radial force balance equation, one finds that this equation gives a small correction term to the toroidal rotation \cite{WangPoP17}.

The parallel momentum "pedestal" due to Eq. (\ref{eq:df}), which represents the effect of boundary trapped ion orbit loss on core passing ion through the topological transition, can be evaluated as
\begin{equation}
\left[nmu\right]_{ped}=\left\langle \int d^3\mathbf{v} mv_{\|}F_{\sigma, ped}\right\rangle \sim -\epsilon_{a}^{3/2}\frac{1}{\Omega_{p}}n\partial_{r}T_{B}, \label{eq:pedestal-m}
\end{equation}
where $\epsilon_{a}=a/R$, with $a$ the minor radius of the torus, $\Omega_{p}=\Omega B_{p}/B$; note that we have used Eq. (\ref{eq:LRW}).
In this estimation, we assumed that the orbit transition takes place in the boundary region where the orbit loss takes place; clearly this gives an underestimate of the intrinsic flow induced by the orbit transition.

Using Eq. (\ref{eq:nu}) and Eqs. (\ref{eq:F-ped}-\ref{eq:pedestal-m}), one finds that the core plasma momentum is given by
\begin{equation}
\left[nmu\right]_{\psi\leq \psi_{B}}=\left[nmu\right]_{ped}+\left[nmu\right]_{d}\sim -\epsilon_{a}^{3/2}\frac{1}{\Omega_{p}}n\partial_{r}T_{B}. \label{eq:core-m}
\end{equation}

Note that this is consistent with Eq. (\ref{eq:anis}).
The pedestal structure of parallel momentum is schematically shown in Fig. 6, which is not hard to understand by examining Eq. (\ref{eq:core-m}), or Fig. 5.

\begin{figure}
  \includegraphics[width=6.0cm]{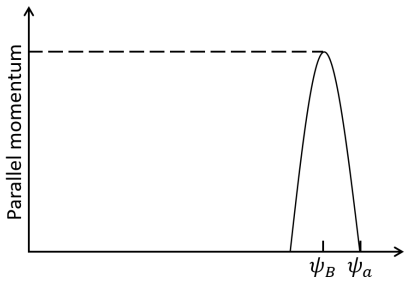}\\
  \caption{Parallel momentum predicted by the present theory (dashed line), in comparison to the previous result (solid line) \cite{PanNF14,StacyNF13,BoedoPoP16}.} \label{fig6}
\end{figure}

For the DIII-D experimental observation \cite{SolomonNF09} of the anomalous co-current momentum source at the edge of the Deuteron H-mode plasma. The main parameters are $R/a=1.6m/0.6m$; $n\sim 5\times10^{19}/m^3$; the poloidal magnetic field at the edge is $\sim 0.15Tesla$. With these parameters, following the above analysis, $T_{B}\sim1keV$ is estimated. The intrinsic parallel flow estimated by using the above theory and parameters is $70km/s$, which agrees well with the experimental observations \cite{SolomonNF09}.
Eq. (\ref{eq:core-m}) predicts a scaling of the intrinsic parallel flow of the core plasma
\begin{equation}
u\sim\epsilon^{3/2}T/I_{p},
\end{equation}
with $I_p$ the plasma current. This is consistent with the well-known Rice-scaling \cite{RiceNF07} of intrinsic toroidal rotation of core plasma,
$u\sim p/I_{p}$.
For a typical ITER plasma, $R/a=6.2m/2m$, $B_T=5.3Tesla$, $B_P=0.3Tesla$ at the edge. $T_{B}\sim3keV$ may be estimated. Eq. (\ref{eq:core-m}) predicts an intrinsic core parallel flow of $\sim100km/s$.

\section{Conclusions and Discussions}
In conclusion, by solving the transport equation with the ion orbit loss boundary condition, we have proposed a kinetic theory of spontaneous core parallel flow induced by turbulence-driven topological transition of orbits in a tokamak H-mode plasma. The proposed theory is consistent with the well-known Rice-scaling \cite{RiceNF07} of intrinsic core plasma flow; it predicts a $\sim100km/s$ intrinsic parallel flow for a typical ITER core plasma. The key point of the proposed mechanism of spontaneous core plasma spin-up in an H-mode tokamak is as follows. Although the orbit loss is dominated by the trapped particles in the boundary region, it affects the distribution of passing particles, which carry the parallel momentum, in the core region; when a co-passing orbit or a counter-passing orbit diffuse to the boundary region, it may change to a trapped orbit through the topological transition. In this way, an asymmetric confinement of passing ions is induced, the confinement region of the co-passing ions is larger than the counter-passing ions, therefore, a spontaneous co-current toroidal flow is maintained. It should be pointed out that a simple orbit-loss model has been adopted in this paper, which assumes that the passing ion orbit-loss does not introduce any significant asymmetry; in a specific machine, the passing ion orbit-loss should be carefully considered, which may mathematically complicate the prediction of the intrinsic flow generated by the proposed mechanism; however, the method developed here can be straightforwardly extended. 

Note that the effect of the residual Reynolds stress \cite{DiamondPoP08,DiamondNF13} depends on the $k_{\|}-$ symmetry-breaking induced by radial electric field. It is shown in Ref. \onlinecite{GriersonPRL17} that the effect of the residual Reynolds stress is radial momentum redistribution, namely, in a system without momentum source, it generates positive momentum in some region with negative momentum in the near region, therefore, it appears that it is not likely to generate a net momentum. The intrinsic rotation due to the turbulence-driven topological transition of orbit does not depends on the $k_{\|}-$ symmetry-breaking, and it does generate a net momentum.

To end this paper, we make some discussions on the collisional effects in the low-collisionality regime, which have been ignored in this paper. Eq. (\ref{eq:Trans}) ignores the collisional effects, as is similar to Ref. \onlinecite{StoltzfusPRL12}, where the effect of turbulence intensity gradient on the ion momentum generation in the H-mode pedestal region was discussed. In Ref. \onlinecite{StoltzfusPRL12}, it was pointed out that a strict justification of ignoring the collisional effects may requires that $\nu_{ii}$, the ion-ion collision rate, is less than the ion turbulent radial diffusion rate and this may be marginally satisfied in the pedestal region of the present tokamaks; it was also pointed out there that the direct collisional effect is not important in ion momentum transport, which is clearly due to the fact that the collision operator conserves the momentum.
A formal treatment to include the collisional effects in the low-collisionality regime can be carried out by slightly extending the methods \cite{BernsteinPF83,ZaitsevPFB93,WangPoP99} previously developed for the Lagrangian formulation of neoclassical transport theory, however, it does not modify the main physics of this paper; the mathematical manipulation is lengthy but straightforward, which is briefly summarized in the following.

When including the collision term $\mathcal{C}\left(f\right)\sim \nu_{ii}f$ in the right-hand side of Eq. (\ref{eq:Trans}), and assuming $\nu_{ii}\tau_{\theta}\sim \delta$, one finds that Eq. (\ref{eq:Trans-Fa}) is modified to

\begin{equation}
\partial_{t} F -\frac{1}{\tau_{\theta}}\partial_{P}\left[\tau_{\theta} \mathcal{D}\left(\overline{\psi}\right)\partial_{P} F\right]-S\left(\overline{\psi}\right)=\bar{\mathcal{C}}\left(F\right). \label{eq:Trans-Fa-C}
\end{equation}

Following Refs. \onlinecite{BernsteinPF83,ZaitsevPFB93,WangPoP99}, one can write the orbital-averaged collisional operator as a divergence of flow in the COM space of $\left(P,\mu,w\right)$. Following Refs. \onlinecite{BernsteinPF83,WangPoP99}, one may sperate the orbital-averaged collision term as
\begin{equation}
\bar{\mathcal{C}}\left(F\right)=\bar{\mathcal{C}}_{0}\left(F\right)+\bar{\mathcal{C}}_{1}\left(F\right)+\bar{\mathcal{C}}_{2}\left(F\right),
\end{equation}
with $\bar{\mathcal{C}}_{1}\left(F\right)$ proportional to $\partial_{P}F$ and $\bar{\mathcal{C}}_{2}\left(F\right)$ proportional to $\partial_{P}\left(...\partial_{P}F\right)$; $\bar{\mathcal{C}}_{0}\left(F\right)$ is independent of $\partial_{P}F$, it only depends on $\partial_{\mu}F$ or $\partial_{w}F$.
Note that \cite{BernsteinPF83,WangPoP99} $\bar{\mathcal{C}}_{1}/\bar{\mathcal{C}}_{0}\sim \mathcal{O}\left(\epsilon_{\rho}\right)$, and $\bar{\mathcal{C}}_{2}/\bar{\mathcal{C}}_{0}\sim \mathcal{O}\left(\epsilon_{\rho}^{2}\right)$, with
\begin{equation}
\epsilon_{\rho}=\rho_{\|}I/\psi \ll1.
\end{equation}

Following Ref. \onlinecite{WangPoP99}, where the equation splitting method \cite{ZwillingerBook97} was used, one may separate Eq. (\ref{eq:Trans-Fa-C}) into
\begin{equation}
\partial_{t} F -\frac{1}{\tau_{\theta}}\partial_{P}\left[\tau_{\theta} \mathcal{D}\left(\overline{\psi}\right)\partial_{P} F\right]-S\left(\overline{\psi}\right)=\bar{\mathcal{C}}_{1}\left(F\right)+\bar{\mathcal{C}}_{2}\left(F\right), \label{eq:Trans-Fa-C1}
\end{equation}
\begin{equation}
\bar{\mathcal{C}}_{0}\left(F\right)=0; \label{eq:C0}
\end{equation}
note that Eq. (\ref{eq:C0}) is similar to Eq. (13b) in Ref. \onlinecite{WangPoP99}.
Following Ref. \onlinecite{BernsteinPF83, WangPoP99}, one finds that Eq. (\ref{eq:C0}) demands that $F$ must be a canonical Maxwellian distribution, $F=F_{M}\left(w;n,T\right)$, with the density and temperature defined on the drift surface: $n=n\left(P,\sigma\right)$, and $T=T\left(P\right)$.

Note that
\begin{equation}
\bar{\mathcal{C}}_{2}\left(F\right)=\frac{1}{\tau_{\theta}}\partial_{P}\left[\tau_{\theta} \mathcal{D}_{neo}\partial_{P} F\right], \label{eq:C1}
\end{equation}
with $\mathcal{D}_{neo}$ the neoclassical radial diffusivity \cite{WangPoP99}.
The neoclassical ion radial diffusivity, $\mathcal{D}_{neo}\sim \epsilon_{\rho}^2 \nu_{ii}$, is usually less than the turbulent ion radial diffusivity, $\mathcal{D}$, therefore one can ignore $\bar{\mathcal{C}}_{2}\left(F\right)$ in Eq. (\ref{eq:Trans-Fa-C1}).

Note that $\bar{\mathcal{C}}_{1}\sim \epsilon_{\rho}\nu_{ii}$. It should be pointed out that for trapped particles, $\bar{\mathcal{C}}_{1}=0$, which is due to the facts \cite{BernsteinPF83,WangPoP99} that $\bar{\mathcal{C}}_{1}\propto \langle v_{\|} \rangle$ and that $\langle v_{\|} \rangle=0$ for trapped particles. By assuming that
\begin{equation}
\epsilon_{\rho}\nu_{ii}<1/\tau_{E},
\end{equation}
one may ignore $\bar{\mathcal{C}}_{1}\left(F\right)$ in Eq. (\ref{eq:Trans-Fa-C1}), which is physically consistent with the momentum conservation of the collision term; note that $\epsilon_{\rho}\sim 0.01$ for the present tokamaks. Therefore, one finds that Eq. (\ref{eq:Trans-Fa-C1}) is reduced to Eq. (\ref{eq:Trans-Fa}). 

In Ref. \onlinecite{McDevittPRL13}, the anomalous pitch-angle scattering of electrons near the TPB induced by the electron radial turbulent diffusion driven by the trapped electron mode, has been numerically demonstrated to be able to drive a bootstrap current; it has been shown there the result in the limit of collisionless plasma is still a reasonable estimate of the typical tokamak fusion plasma in the low-collisionality regime.

Neglecting the effects of ion-ion collision in the low-collisionality regime in this paper can also be understood in the following simple way. Introduce a modified Krook operator,
\begin{equation}
\mathcal{C}\left(f\right)\simeq-\nu_{ii}\left(f-f_{M}\right),
\end{equation}
with $f_{M}$ a local shift-Maxwellian distribution defined on the magnetic flux surface, and $\nu_{ii}$ the ion-ion collision rate. The point is that when the ion distribution is a shift-Maxwellian, the ion-ion collision term is zero. Therefore, ignoring the collision term does not need the too strict requirement that the collision rate is less than the radial turbulent diffusion rate; it can be more easily justified by requiring that the distribution function is not too much deviated from a shift Maxwellian distribution. Since it is the finite banana-width effect that is discussed here, one may estimate the deviation of the distribution function induced by this effect as $\left(F-f_{M}\right)/f_{M}\sim \Delta \psi_{B}/\psi_a$ [see, Eq. (\ref{eq:anis})], this number is typically $0.05$ for the present tokamak plasmas; using this number, one finds that $\mathcal{C}\left(f\right)<f/\tau_{E}$ can be roughly satisfied in many tokamaks at the present. We point out here that by ignoring the collision term when discussing physics on a time scale longer than the collision time, the final results should not depend on a strongly non-Maxwellian distribution. It should also be pointed out that many nonlinear gyrokinetic simulations ignore the collisional effects, therefore, the present collisionless model is at least useful in understanding the relevant collisionless gyrokinetic simulation results.

Additional Information: The author declares no competing interests.

\begin{acknowledgments}
The author is in debt to S. Sun for his help in preparing Figs. 2-4, which are produced by the NLT nonlinear ITG turbulence simulation data. This work was supported by the National Natural Science Foundation of China under Grant Nos. 11875254 and 11675176.
\end{acknowledgments}

\nocite{*}

\end{document}